# MERGERS OF GLOBULAR CLUSTERS

SIDNEY VAN DEN BERGH

Dominion Astrophysical Observatory

5071 West Saanich Road

Victoria, British Columbia

V8X 4M6, Canada

vandenbergh@dao.nrc.ca

## ABSTRACT

Globular clusters with composite color-magnitude diagrams, such as NGC 1851, NGC 2808 and Fornax No. 3, might have formed by mergers. It is suggested that each of these objects had two parent clusters, one with a red horizontal branch and another with a blue horizontal branch. Such mergers could have occurred if both ancestral objects were originally members of a single dwarf spheroidal galaxy in which the internal velocity dispersion was low.

## 1. INTRODUCTION

Globular clusters, that encounter each other at low relative velocities, will lose angular momentum and merge into a single cluster on a relatively short time-scale (Sugimoto & Makino 1989). The spiraling in of their orbits is due to (1) loss of angular momentum which is carried away by escaping stars, and (2) to synchronization of the spin with orbital rotation which proceeds at the expense of orbital angular momentum. Merger calculations for ancestral globular clusters with unequal masses have been published by Makino, Akiyama & Sugimoto (1991) who find that the end products of mergers have ellipticities in the range 0.25 to 0.35. (However, the merger product may appear more circular when viewed in projection.) Mergers between globular clusters are expected to be qualitatively similar to those between spherical galaxies which have been modeled by White (1978), except that the former are usually considered to have bound elliptical orbits, whereas unbound parabolic or hyperbolic orbits are used for the latter. From his N-body simulations White concluded that galaxies formed by merger of spherical ancestors generally tend to be flattened towards their orbital plane. Furthermore, the merger products were found to have compact cores and more extended envelopes than their progenitors. However, in the case of globular clusters such extended envelopes might be trimmed back by Galactic tidal forces.

Ancestral globular clusters that merged could have had differing compositions, ages or evolutionary histories. Their merger product might therefore exhibit a "composite" color magnitude diagram. Perhaps the best-known example of such a globular cluster with a composite horizontal branch is NGC 2808 (Ferrarro et al. 1990). Rood et al. (1993) show that the morphology of the horizontal branch (HB) of this cluster resembles that of a super-position of the HBs of NGC 288 and NGC 362. These clusters are examples of objects that exhibit "second parameter" (van den Bergh 1965, 1967, Sandage & Wildey 1967) effects in opposite directions.

It has been argued (Rood et al. 1993) that the existence of clusters having both a red HB clump <u>and</u> an extended blue branch proves that the second parameter cannot be exclusively an age effect. This is so because, at a given metallicity, clusters with red HBs are a few Gyr younger than globulars with blue HBs. Alternatively, one could make the <u>ad hoc</u> assumption that stars on the horizontal branches of objects like NGC 2808 suffered bimodal mass loss.

Other clusters that are known to exhibit composite color-magnitude diagrams are NGC 1851 (Walker 1992) and Cluster No. 3 in the Fornax dwarf spheroidal galaxy (Buonanno et al. 1996).

## 2.  DISCUSSION

In this <u>letter</u> it is proposed that globular clusters with composite HBs might have been formed by mergers between one parent cluster that had a red HB and another cluster with a blue HB.  For such mergers to occur individual clusters must have quite small relative velocities.  Low relative velocities will be exceedingly rare between members of the Galactic globular cluster system, which has a velocity dispersion $\sim 10^2$ km s$^{-1}$.  However, such mergers could occur in dwarf spheroidal galaxies which are observed to have central velocity dispersions in the range 6.5 - 11 km s$^{-1}$ (Mateo 1996).  In fact, Cluster No. 3 in the Fornax dwarf spheroidal galaxy (Buonanno et al. 1996), which appears to have a composite color-magnitude diagram, may be an example of a globular cluster in a dwarf spheroidal galaxy that has undergone such a merger.

Fornax clusters Nos. 1 and 2 have radial density profiles (Rodgers & Roberts 1994) that appear to fit truncated King models.  Rodgers & Roberts speculate that this truncation may be due to past dynamical interactions.  With slightly different interaction parameters these two clusters might also have merged!  However, the fact that the horizontal branch of cluster No. 1 has an intermediate color (Buonanno et al. 1996) implies that it would have been difficult to recognize the merger remnant as such from its color-magnitude diagram.  One might also speculate that the exceptionally bright ($M_V = -9.9$) cluster M54 (= NGC 6715), which is located near the center of the Sagittarius dwarf, owes its high luminosity to a merger. The color-magnitude diagrams (Sarajedini & Layden 1996) of the clusters Arp 2 (blue) and Terzan 7 (red), which both seem to be associated with the Sagittarius dwarf (Ibata, Gilmore & Irwin 1994), are also excellent examples of clusters that could have produced a composite color-magnitude diagram if they had merged.

Possibly NGC 1851 and NGC 2808 were once located in dwarf spheroidals that were subsequently destroyed by Galactic tidal forces (NGC 1851 has $R_{GC}$ = 16.3 kpc, and NGC 2808 has $R_{GC}$ = 10.7 kpc). The search for remnants of such a dwarf spheroidal would be easier in the case of NGC 1851 (b = -35°) than for NGC 2808 (b = -11°). *If* age is indeed the second parameter then the red HB parent of each of the clusters with composite horizontal branches must be a few Gyr younger than the parent with a blue horizontal branch. This requirement is not inconsistent with the observation that star formation seems to have continued for many Gyr in some dwarf spheroidal (e.g. Mighell & Rich 1996). However, Walker's (1992) observations of main sequence stars in NGC 1851 appear to militate against the merger hypothesis. The small [$\Delta$ (B-V) ~ 0.05 mag] intrinsic width of the NGC 1851 main-sequence argues against large metallicity differences between individual stars in this cluster. Furthermore the lack of structure in the cluster color-magnitude diagram near the main-sequence turnoff appears to rule out age differences of more than a couple of Gyr among cluster members. Alternatively, it might be argued that the absence of large age differences in NGC 1851 militates against age being the primary cause of second parameter effects (Stetson, VandenBerg & Bolte 1996). The fact that neither NGC 1851 ($\epsilon = 0.02$) nor NGC 2808 ($\epsilon = 0.13$) have unusually large ellipticities (Frenk & Fall 1982) does not favor the merger hypothesis. However, an argument in favor of the idea that NGC 1851 was involved in a merger is provided by its unusually high central concentration of light. The N-body calculations of White (1978) showed that the final object formed by merging two initially spherical star clusters is more centrally concentrated than its ancestors. In this connection, it is of interest to note that NGC 1851 is one of the most centrally concentrated globular clusters in the outer halo of the Galaxy. According to Trager, Djorgovski & King (1993) it has $C = \log (r_t / r_c) = 2.24$. Only three of 50 (6%) of all clusters with $R_{GC} > 10$ kpc listed by Trager et al. have $C > 2.0$. [The two other very compact outer halo globular clusters are NGC 5824 ($C = 2.45$) and NGC 7078 ($C = 2.50$)]. The other merger suspect NGC 2808 is also quite compact. Its value $C = 1.77$ places it among the 22% of all outer halo clusters with $C > 1.75$. It is concluded that the high central concentration of light in NGC 1851 and NGC 2808 gives some encouragement to the notion that these clusters might represent end products of mergers.

Both NGC 1851 and NGC 2808 belong to the α population of van den Bergh (1993). NGC 1851 appears to be on a retrograde orbit, whereas NGC 2808 has prograde motion. It is also noted in passing that NGC 1851 is located near the Fornax-Leo-Sculptor (F-L-S) plane of Fusi Pecci et al. (1995), and that NGC 2808 lies near the intersection of the Magellanic plane and the F-L-S plane.

It is tentatively concluded that some, but not all, presently available data are consistent with the hypothesis that a few Galactic globular clusters with composite color-magnitude diagrams may have been formed by mergers between parental globulars that were once located within dwarf spheroidal galaxies.

It is a pleasure to thank a particularly helpful anonymous referee. I also thank Piet Hut and Eric LeBlanc for providing a number of useful references, and Peter Stetson for stimulating discussions.